\documentclass[twocolumn,showpacs,superscriptaddress,amsmath,amssymb]{revtex4}

\usepackage{amsmath}
\usepackage{graphicx}
\usepackage{dcolumn}

\begin{document}

\title{Nuclear Structure Calculations and Modern 
Nucleon-Nucleon Potentials}
\author{L. Coraggio}
\affiliation{Dipartimento di Scienze Fisiche, Universit\`a
di Napoli Federico II, \\ and Istituto Nazionale di Fisica Nucleare, \\
Complesso Universitario di Monte  S. Angelo, Via Cintia - I-80126 Napoli,
Italy}
\author{A. Covello}
\affiliation{Dipartimento di Scienze Fisiche, Universit\`a
di Napoli Federico II, \\ and Istituto Nazionale di Fisica Nucleare, \\
Complesso Universitario di Monte  S. Angelo, Via Cintia - I-80126 Napoli,
Italy}
\author{A. Gargano}
\affiliation{Dipartimento di Scienze Fisiche, Universit\`a
di Napoli Federico II, \\ and Istituto Nazionale di Fisica Nucleare, \\
Complesso Universitario di Monte  S. Angelo, Via Cintia - I-80126 Napoli,
Italy}
\author{N. Itaco}
\affiliation{Dipartimento di Scienze Fisiche, Universit\`a
di Napoli Federico II, \\ and Istituto Nazionale di Fisica Nucleare, \\
Complesso Universitario di Monte  S. Angelo, Via Cintia - I-80126 Napoli,
Italy}
\author{T. T. S. Kuo}
\affiliation{Department of Physics, SUNY, Stony Brook, New York 11794}
\author{R. Machleidt}
\affiliation{Department of Physics, University of Idaho, Moscow, Idaho 83844}

\date{\today}

\begin{abstract}
We study ground-state properties of the doubly magic nuclei $^4$He, 
$^{16}$O, and $^{40}$Ca employing the Goldstone expansion and using
as input four different high-quality nucleon-nucleon ($NN$) potentials. 
The short-range repulsion of these potentials is renormalized by 
constructing a smooth low-momentum potential $V_{\rm low-k}$. 
This is used directly in a Hartree-Fock approach and corrections up to 
third order in the Goldstone expansion are evaluated. 
Comparison of the results shows that they are only slightly dependent 
on the choice of the $NN$ potential.
\end{abstract}

\pacs{21.30.Fe, 21.60.Jz, 21.10.Dr}

\maketitle

\section{Introduction}
In recent years, the study of the properties of nuclear systems
starting from a free nucleon-nucleon ($NN$) potential $V_{NN}$ has 
become a subject of special interest.
This has been stimulated by the substantial progress made during the last 
decade in the development of $NN$ potentials that reproduce with high precision
the $NN$ scattering data and deuteron properties 
\cite{stoks94,wiringa95,cdbonn01,entem03}.
However, the fact that these potentials predict almost identical 
phase shifts does not imply, owing to their different off-shell
behavior, that they should give the same results when employed in 
nuclear many-body calculations. 
It is therefore of great interest to investigate how much nuclear 
structure results depend on the $NN$ potential one starts with, namely 
to try to assess the relevance of the off-shell effects in microscopic 
nuclear structure calculations.

The differences between various $NN$ potentials in describing
properties of nuclear matter have been investigated by several
authors, the main aim being to try to assess the role of the various 
components of the nuclear force.
In this context, we may mention the studies of Refs. 
\cite{engvik97,baldo98,muther99,frick02}, where attention has been 
focused on modern phase-shift equivalent $NN$ potentials. 

As regards the study of finite nuclei, while a rather large number of
realistic nuclear structure calculations have been carried out in the 
past few years,  only a few attempts have been made 
\cite{jiang92,hjorth95,covello99} to study to which extent these 
calculations depend on the $NN$ potential used as input. 
Actually, no detailed comparison of the results produced by different 
phase-shift equivalent potentials in the description of nuclear
structure properties has yet been done. 
It may be worth mentioning, however, the work of
Ref. \cite{machleidt96}, where different high-precision $NN$
potentials have been considered to investigate the effects of 
nonlocalities on the triton binding energy. 

On the above grounds, we have found it interesting and timely to 
perform nuclear structure calculations using different phase-shift 
equivalent $NN$ potentials and make a detailed comparison between the 
corresponding results.
In a recent paper \cite{coraggio03}, we performed realistic
calculations of the ground-state properties of some doubly magic
nuclei within the framework of the Goldstone expansion approach, and 
showed that the rate of convergence is very satisfactory.
Motivated by the results obtained in that work, in the present paper 
we make use of the Goldstone expansion to calculate the binding
energies and rms charge radii of $^4$He, $^{16}$O, and $^{40}$Ca 
for different phase-shift equivalent $NN$ potentials. 
We feel that this is a good ``laboratory'' for a comparative study 
of the effects of $NN$ potentials in finite nuclei. 
We consider the four high-quality $NN$ potentials Nijmegen II 
\cite{stoks94}, Argonne V18 \cite{wiringa95}, CD-Bonn \cite{cdbonn01}, 
and N$^3$LO \cite{entem03}.

As is well known, to perform nuclear structure calculations
with realistic $NN$ potentials one has to deal with the strong
repulsive behavior of such potentials in the high-momentum regime.
Recently, a new method to renormalize the bare $NN$ interaction 
has been proposed \cite{bogner01,bogner02}, which is proving to be an 
advantageous alternative to the use of the Brueckner $G$ matrix 
\cite{bogner01,coraggio02a,coraggio02b,covello02}.
It consists in deriving from $V_{NN}$ a low-momentum potential 
$V_{\rm low-k}$ defined within a cutoff momentum $\Lambda$. 
This is a smooth potential which preserves exactly the on-shell
properties of the original $V_{NN}$ and is suitable for being used 
directly in nuclear structure calculations. 

As in our earlier work \cite{coraggio03}, we construct the $V_{\rm
low-k}$ for each of the four above mentioned $NN$ potentials.
The various $V_{\rm low-k}$'s are then used directly in Hartree-Fock 
calculations.  
Once the self-consistent basis is obtained, we calculate the Goldstone
expansion including diagrams up to third order in $V_{\rm low-k}$.

It is worth emphasizing that one of the main advantages of the $V_{\rm
low-k}$ renormalization method, with respect to the $G$-matrix one, is
to preserve the phase-shift equivalence of the original $NN$
potentials.
Thus, it is particularly interesting to compare the results obtained
in nuclear structure calculations employing on-shell equivalent 
$V_{\rm low-k}$'s. 

The paper is organized as follows. 
In Sec. II we give a brief description of the main features of the
four phase-shift equivalent $NN$ potentials considered in our study.
In Sec. III we give an outline of the derivation of $V_{\rm low-k}$ 
and some details of our calculations. 
In Sec. IV we present and discuss our results. 
Some concluding remarks are given in Sec. V.

\section{Realistic Nucleon-Nucleon Potentials} 
As reported in the Introduction, we employ the Nijmegen II \cite{stoks94}, 
Argonne V18 \cite{wiringa95}, CD-Bonn \cite{cdbonn01}, and chiral 
N$^3$LO \cite{entem03} $NN$ potentials, which have all been fitted to 
the Nijmegen phase-shift analysis as well as the proton-proton and 
neutron-proton data below 350 MeV \cite{stoks93}.
It is well known that these potentials, even if they reproduce the
$NN$ data with almost the same accuracy, may have a rather different 
mathematical structure.
The Nijmegen II and the Argonne V18 potentials are non-relativistic
and defined in terms of local functions, which are multiplied by
a set of spin, isospin and angular momentum operators.
The CD-Bonn potential, based on relativistic meson field theory, is 
represented in terms of the covariant Feynmann amplitudes for
one-boson exchange, which are nonlocal \cite{machleidt89}.
The N$^3$LO potential is based upon a chiral effective Lagrangian.
The model includes one- and two-pion exchange contributions
and so-called contact terms up to chiral order four, some of which are 
non-local.

\noindent
All the above $NN$ interactions reproduce equally well the same
phase-shifts, so the corresponding on-shell matrix elements of the 
reaction matrix $T$ are the same as well.
However, this does not imply that the interactions are identical.
The $T$-matrix is obtained from the Lippmann-Schwinger equation

\[ T(k',k,k^2) = V_{NN}(k',k) + ~~~~~~~~~~~~~~~~~~~~~~~~~~~~~~~~~~~~~~~\]
\begin{equation}
~~~~~~~~~~~~~~+{\cal P}\int _0 ^{\infty} q^2 dq  V_{NN}(k',q)
\frac{1}{k^2-q^2} T(q,k,k^2 ) ~~,
\end{equation}

\noindent
where $k,~k'$, and $q$ stand for the relative momenta.
Notice that the $T$-matrix is the sum of two terms, the Born term and
an integral term.
Notwithstanding the sum is the same, the individual terms may still be
quite different.
For example, in Fig.~1 we show the value of the $^3{\rm S}_1$ $T$-matrix 
elements for $K_{\rm lab}=150$ MeV and $k=k'=k_0=1.34 \; {\rm fm}^{-1}$.
They are indicated by the full circle and are practically identical
for the four $NN$ potentials we have employed.
In Fig. 1 we display also the matrix elements $V_{NN}(k_0,k)$ as a
function of $k$ for the four potentials.
The asterisks stand for the diagonal matrix elements
$V_{NN}(k_0,k_0)$, which represent the Born approximation to $T$.
From the inspection of this figure, it is clear that even if different
$V_{NN}$'s reproduce the same $T$-matrix element, the latter is obtained
by summing two terms that are significantly different for each
potential.
More precisely, it is evident that Nijmegen II potential is a quite
``hard'' potential, its diagonal matrix element being very repulsive.
So, in order to reproduce correctly the on-shell $T$-matrix element,
it needs a large attractive contribution from the integral term
of Eq. (1), the latter being related to the tensor component of the $NN$
force and to its off-shell behavior (see Ref. \cite{machleidt94} for a
closer examination).
On the other hand, the N$^3$LO interaction is a rather ``soft'' 
potential, implying a smaller contribution from the integral term. 

\begin{figure}[ht]
\includegraphics[scale=0.8,angle=0]{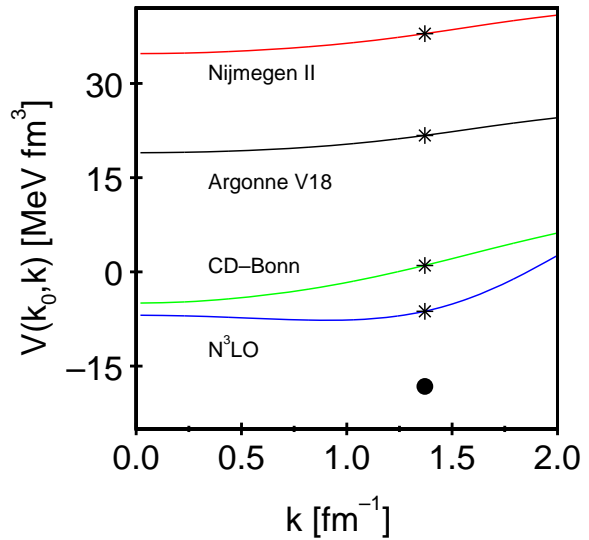}
\caption{Matrix elements $V_{NN}(k_0,k)$  for the $^3 {\rm S}_1$ 
partial wave for CD-Bonn (green line), Nijmegen II (red line), 
Argonne V18 (black line), and N$^3$LO (blue line) potentials. 
The diagonal matrix elements $k_0=1.34~ {\rm fm}^{-1}$ are marked by an
asterisk. 
The corresponding $T$-matrix element is marked by a full circle (see
text for more details).}
\end{figure}

\noindent
Similar features may be observed also in the other partial waves, 
however,
the differences among the potentials decrease for larger values of 
the orbital angular momentum.

\section{Method of calculation}
A traditional approach to the renormalization of the strong repulsive
behavior of realistic $NN$ potentials, when dealing with doubly
closed-shell nuclei, is the Brueckner-Goldstone (BG) theory (see for 
instance Refs. \cite{Day67,towner}), where the Goldstone perturbative 
expansion is re-ordered summing to all orders only the ladder diagrams. 
Consequently, the bare interaction ($V_{NN}$) vertices are replaced 
by the reaction matrix ($G$).
This framework leads to the well known Brueckner-Hartree-Fock 
(BHF) theory, when the self-consistent definition is adopted for the
single-particle (SP) auxiliary potential and only the first-order
contribution in the BG expansion is taken into account. 
So, the BHF approximation gives a mean field description of the 
ground state of nuclei in terms of the $G$ matrix, the latter taking 
into account the correlations between pairs of nucleons. 
However this procedure is not without difficulties, because of the 
energy dependence of $G$.

As already mentioned in the Introduction, we renormalize the short-range 
repulsion of the bare $NN$ potential by integrating out its high
momentum components \cite{bogner01,bogner02}. 
The resulting low-momentum potential, $V_{\rm low-k}$, is a smooth 
potential, whose vertices can be used directly to sum up the
Goldstone expansion diagrams.

According to the general definition of a renormalization group
transformation, $V_{\rm low-k}$ must be such that the low-energy 
observables calculated in the full theory are preserved exactly by the 
effective theory.

For the nucleon-nucleon problem in vacuum, we require that the deuteron
binding energy, low-energy phase shifts, and low-momentum
half-on-shell $T$ matrix calculated from $V_{NN}$ must be reproduced
by $V_{\rm low-k}$.
The effective low-momentum T matrix is defined by

\[T_{\rm low-k }(p',p,p^2) = V_{\rm low-k }(p',p) + ~~~~~~~~~~~~~~~~~~~~~~~~~~~~~~~~~~~~~~~\]
\begin{equation}
~~~~+{\cal P}\int _0 ^{\Lambda} q^2 dq
V_{\rm low-k }(p',q) \frac{1}{p^2-q^2} T_{\rm low-k} (q,p,p^2)~~.
\end{equation}

\noindent
Note that for $T_{\rm low-k }$ the intermediate states are integrated up
to $\Lambda$.

It is required that, for $p$ and $p'$ both belonging to $P$ ($p,p' \leq
\Lambda$), $T(p',p,p^2)= T_{\rm low-k }(p',p,p^2)$.
In Refs. \cite{bogner01,bogner02} it has been shown that the above
requirements are satisfied when $V_{\rm low-k}$ is given by the
folded-diagram series

\[ V_{\rm low-k} = \hat{Q} - \hat{Q'} \int \hat{Q} + \hat{Q'} \int \hat{Q}
\int \hat{Q} -~~~~~~~~~~~~~~~~~\] 
\begin{equation}
~~~~~-\hat{Q'} \int \hat{Q} \int \hat{Q} \int \hat{Q} + ...~~,
\end{equation}

\noindent
where $\hat{Q}$ is an irreducible vertex function, in the sense that
its intermediate states must be outside the model space $P$.
The integral sign represents a generalized folding operation \cite{krenc80},
and $\hat{Q'}$ is obtained from $\hat{Q}$ by removing terms of first
order in the interaction.

The above $V_{\rm low-k}$ can be calculated by means of iterative
techniques.
We have used here an iteration method proposed in Ref. \cite{andre96}, 
which is particularly suitable for non-degenerate model spaces.
This method, which we refer to as Andreozzi-Lee-Suzuki (ALS) method, 
is an iterative method of the Lee-Suzuki type \cite{suzuki80}.

To exemplify how $V_{\rm low-k}$ preserves low-energy observables,
we report in Table I the deuteron binding energy, and the
neutron-proton $^1S_0$ phase shifts calculated both with the full
CD-Bonn potential and its $V_{\rm low-k}$ (with a cut-off momentum
$\Lambda=2.0~{\rm fm}^{-1}$).

\begin{table}[ht]
\caption{Deuteron binding energy (MeV) and $np$ $^1S_0$ phase shifts
(deg) as predicted by full CD-Bonn and its $V_{\rm low-k}$ 
($\Lambda=2.0~{\rm fm}^{-1}$)}
\begin{ruledtabular}
\begin{tabular}{llccc}
  & & CD-Bonn & $V_{\rm low-k}$ & Expt. \\
\colrule
$B_d$ &  & 2.224 & 2.224 & 2.224 \\
\colrule
Phase shifts &   &  &  &  \\ 
 & $E_{\rm lab}$ & & & \\
             & 1   & 62.1 & 62.1 & 62.1 \\ 
             & 10  & 60.0 & 60.0 & 60.0 \\ 
             & 25  & 50.9 & 50.9 & 50.9 \\ 
             & 50  & 40.5 & 40.5 & 40.5 \\ 
             & 100 & 26.4 & 26.4 & 26.8 \\ 
             & 150 & 16.3 & 16.3 & 16.9 \\ 
             & 200 & 8.3  & 8.3  & 8.9  \\ 
             & 250 & 1.6  & 1.6  & 2.0  \\ 
             & 300 & -4.3 & -4.3 & -4.5 \\ 
\end{tabular}
\end{ruledtabular}
\end{table}

An important question in this approach is what value one should use for 
the cutoff momentum. 
A discussion of this point as well as a criterion for the choice
of $\Lambda$ can be found in Ref. \cite{bogner02}. 
According to this criterion, we have used here $\Lambda=2.1 \; 
{\rm fm}^{-1}$.

After having renormalized the various $NN$ potentials, we use
the corresponding $V_{\rm low-k}$'s directly in a HF calculation. 
The HF equations are then solved for $^{4}$He, $^{16}$O and $^{40}$Ca 
making use of a harmonic-oscillator (HO) basis.
The details of the HF procedure are reported in Ref. \cite{coraggio03}.
As a major improvement, in this work we remove the spurious
center-of-mass kinetic energy writing the kinetic energy operator $T$ as

\begin{equation}
T= \frac{1}{2Am} \sum_{i<j} ({\rm {\bf p}}_i - {\rm {\bf p}}_j )^2 ~~.
\end{equation}

Similarly, we define the mean square radius operator as

\begin{equation}
r^2= \frac{1}{A^2} \sum_{i<j} ({\rm {\bf r}}_i - {\rm {\bf r}}_j )^2 ~~.
\end{equation}

A complete review about center-of-mass correction in self-consistent
theories may be found in Ref. \cite{davies71}

In our calculations the HF SP states are expanded in a finite series
of $N=5$ harmonic-oscillator wave-functions for $^{16}$O and
$^{40}$Ca, and $N=6$ for $^{4}$He.
This truncation is sufficient to ensure that the HF results do not 
significantly depend on the variation of the oscillator constant 
$\hbar \omega$, as we showed in Ref. \cite{coraggio03}. 
The values  of $\hbar \omega$ adopted here have been derived from
the expression $\hbar \omega= 45 A^{-1/3} -25 A^{-2/3}$ \cite{blomqvist68},
which reproduces the rms radii in an independent-particle approximation
with harmonic-oscillator wave functions.
This expression gives  $\hbar \omega= 18$, 14 and 11 MeV for $^{4}$He,
$^{16}$O and $^{40}$Ca, respectively. 

We use the HF basis to sum both the Goldstone expansion and the 
diagrams for the mean square charge radius $\langle r^2 \rangle$, 
including contributions up to third order in $V_{\rm low-k}$. 
Fig. 2 shows first-, second-, and third-order diagrams \cite{goldstone57} 
of the Goldstone expansion.

\begin{figure}[ht]
\includegraphics[scale=0.55,angle=-90]{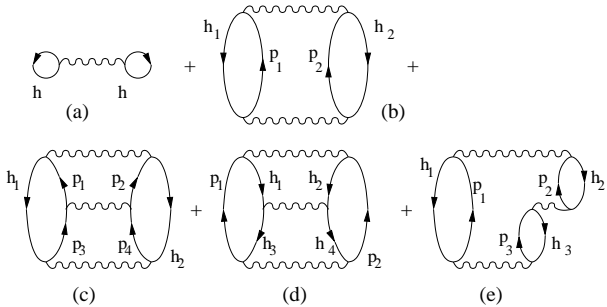}
\caption{First-, second-, and third-order diagrams in the Goldstone
expansion.}
\end{figure}

\section{Results}

In Table II we show for $^4$He, $^{16}$O and $^{40}$Ca the calculated 
binding energy per nucleon and the rms charge radius  obtained using 
different phase-shift equivalent $NN$ potentials, and compare them
with the experimental data \cite{audi93,devries87,nadjakov94}.

\begin{table}[ht]
\caption{Comparison of the calculated binding energies per nucleon 
(MeV/nucleon) and rms radii (fm) for different $V_{\rm NN}$ with the 
experimental data for $^{4}$He, $^{16}$O and $^{40}$Ca.
We take into account the finite dimensions of the proton using an 
estimate of its rms charge radius $\sqrt{r^2_p}=0.8$ fm \cite{devries87}.}
\begin{ruledtabular}
\begin{tabular}{llccccc}
Nucleus &  & Nijmegen II & AV18  & CD-Bonn & N3LO & Expt. \\
\colrule
 $^{4}$He & & & & & &\\
 & $BE/A$         & 6.88 & 6.85 & 6.95 & 6.61 & 7.07 \\
 &$\langle r^2 \rangle^{1/2}$& 1.68 & 1.69 & 1.63 & 1.75 & 1.67 \\
 $^{16}$O & & & & & &\\
 & $BE/A$         & 8.26 & 8.26 & 8.30 & 8.11 & 7.98 \\
 &$\langle r^2 \rangle^{1/2}$& 2.58 & 2.59 & 2.49 & 2.66 & 2.73 \\
 $^{40}$Ca & & & & & &\\
 & $BE/A$         & 9.66 & 9.53 & 9.93 & 9.50 & 8.55 \\
 &$\langle r^2 \rangle^{1/2}$& 3.20 & 3.22 & 3.10 & 3.29 & 3.485 \\
\end{tabular}
\end{ruledtabular}
\end{table}

A detailed analysis about the convergence properties of the
perturbative series can be found in Ref. \cite{coraggio03}, where it
is shown that the convergence is fairly rapid and higher-order 
contributions are negligible.

From Table II, we see that the calculated quantities are scarcely 
sensitive to the choice of the $NN$ potential.

\begin{table}[ht]
\caption{Calculated $P_D$'s with different $NN$ potentials and 
with the corresponding $V_{\rm low-k}$'s}
\begin{ruledtabular}
\begin{tabular}{lccccc}
  & Nijmegen II & AV18  & CD-Bonn & N3LO \\
\colrule
 Full potential       & 5.63 & 5.76 & 4.85 & 4.51 \\
 $V_{\rm low-k}$      & 4.32 & 4.37 & 4.04 & 4.32 \\
\end{tabular}
\end{ruledtabular}
\end{table}

As a matter of fact, the binding energies per nucleon and the rms charge 
radii calculated using the various potentials differ at most by 0.43 MeV 
and 0.19 fm, respectively.
This insensitivity may be traced back to the fact that when renormalizing 
the short-range repulsion of the various potentials, the differences 
existing between their off-shell properties are attenuated.
It is well known (see for instance \cite{machleidt94}) that the
off-shell behavior of a potential, and in particular its off-shell 
tensor force strength, is related to the $D$-state probability of 
the deuteron $P_D$; this is why when comparing $NN$ potentials, 
off-shell differences are seen in $P_D$ differences.
For this reason, we report in Table III the predicted $P_D$'s for each of
the potentials under consideration, and compare them with those
calculated with the corresponding $V_{\rm low-k}$'s.
We see that while the $P_D$'s given by the full potentials are
substantially different, ranging from 4.5 to 5.8 \%, they become quite
similar after renormalization. 
This is an indication that the ``on-shell equivalent'' potentials we 
have used are made almost ``off-shell equivalent'' by the
renormalization procedure.
This aspect is more evident when comparing our calculated $^4$He
binding energies with the results of exact calculations, based on the 
Faddeev-Yakubovsky procedure \cite{nogga00,navratil04} (see Table IV).
As a matter of fact, the difference between the exact results, that is at
most 2 MeV, is reduced to 1.3 MeV in our calculations.
It has to be observed that larger differences between our
results and the exact ones are obtained when employing high-$P_D$ 
potentials, such as Nijmegen II and Argonne V18.
Smaller differences occur for the N$^3$LO (1 MeV) and CD-Bonn (1.5
MeV) potentials, whose tensor force strengths are smaller than those
of the two other potentials.
This reflects the fact that for these two potentials the
renormalization procedure modifies the original $P_D$ to a limited
extent, while a stronger change occurs for the Nijmegen II and Argonne 
V18 potentials.
Table IV also shows that in all cases we get more binding than the
exact calculations, as a consequence of the renormalization of the repulsive
components of the potentials.

\begin{table}[ht]
\caption{Comparison of the $^4$He calculated binding energies 
(MeV) for different $V_{\rm NN}$ obtained using Goldstone expansion (I) and 
with Faddeev-Yakubovsky procedure (II).}
\begin{ruledtabular}
\begin{tabular}{lcccc}
 $BE$ & Nijmegen II & AV18  & CD-Bonn & N3LO \\
\colrule
 (I)        & 27.523 & 27.409 & 27.799 & 26.440 \\
 (II)       & 24.560 & 24.280 & 26.260 & 25.410 \\
\end{tabular}
\end{ruledtabular}
\end{table}

\section{Summary and conclusions}
The aim of this work has been to compare the results of microscopic 
nuclear structure calculations, starting from four different
phase-shift equivalent $NN$ potentials, Nijmegen II, Argonne V18, CD-Bonn,
and N$^3$LO.
To this end, we have calculated ground-state properties of the doubly 
closed nuclei $^4$He, $^{16}$O, and $^{40}$Ca by way of the Goldstone 
expansion.
This has been done within the framework of the so-called $V_{\rm low-k}$ 
approach \cite{bogner01,bogner02} to the renormalization of the 
short-range repulsion of the $NN$ potentials, wherein a 
low-momentum potential is derived, which preserves the low-energy physics of 
the original potential.

The analysis of the  results  obtained shows that the calculated
properties are only weakly dependent on the $NN$ potential used as
input. 
This result may be traced back to the renormalization procedure of
the short-range repulsion.
As a matter of fact, we have shown that the renormalized potentials
are characterized by a reduced off-shell tensor force strength,
as compared with that of the original potential.
Moreover, the $P_D$'s of the different potentials become quite
similar, this quantity being related to the balance between the
central and tensor components of the nuclear force.

It is worthwhile to point out that when dealing with the N$^3$LO
chiral potential, the renormalization procedure, which by construction
preserves exactly the on-shell properties up to the cutoff momentum, 
scarcely modifies the off-shell behavior.
This feature, which is related to the fact that chiral perturbation 
theory is a low-momentum expansion, may make this kind of $NN$ 
potentials particularly tailored for microscopic nuclear structure 
calculations.

\begin{acknowledgments}
This work was supported in part by the Italian Ministero
dell'Istruzione, dell'Universit\`a e della Ricerca  (MIUR), by the
U.S. DOE Grant No.~DE-FG02-88ER40388, and by the U.S. NSF Grant 
No.~PHY-0099444.
\end{acknowledgments}

\end{document}